\documentclass{elsart}

\usepackage{graphicx}
\usepackage{bm} 

\def\slashchar#1{\setbox0=\hbox{$#1$}
   \dimen0=\wd0 \setbox1=\hbox{/} \dimen1=\wd1
   \ifdim\dimen0>\dimen1 \rlap{\hbox to \dimen0{\hfil/\hfil}} #1
   \else  \rlap{\hbox to \dimen1{\hfil$#1$\hfil}} / \fi}

\begin{document}

\begin{frontmatter}
  \title{\bf Impact-parameter dependence 
of the generalized parton distribution of the pion in chiral quark models 
\thanksref{grant}} \thanks[grant]{Supported in part by the Spanish
Ministerio de Asuntos Exteriores and the Polish State Committee for
Scientific Research, grant number 07/2001-2002, 
by the Spanish DGI, grant no. BFM2002-03218, and by Junta de
Andaluc\'{\i}a, grant no. FQM-225.}
\thanks[emails]{%
\hspace{0mm} E-mail: b4bronio@cyf-kr.edu.pl, earriola@ugr.es}
\author[INP]{Wojciech Broniowski} and 
\author[Granada]{Enrique Ruiz Arriola} 
\address[INP]{The H. Niewodnicza\'nski Institute of Nuclear Physics,
        ul. Radzikowskiego 152,  PL-31342 Krak\'ow, Poland}
\address[Granada]{Departamento de F\'{\i}sica Moderna, Universidad de
Granada, E-18071 Granada, Spain}

\begin{abstract}
We compute the off-forward diagonal (non-skewed) non-singlet generalized parton
distribution of the pion in two distinct chiral quark models: the
Nambu-Jona-Lasinio model with the Pauli-Villars regulator and the 
Spectral Quark Model. The analysis is carried out in the
impact-parameter space. Leading-order perturbative QCD evolution is
carried out via the inverse Mellin transform in the index space. The
model predictions agree very reasonably with the recent results from
transverse-lattice calculations, reproducing qualitatively both the
Bjorken-$x$ and the impact-parameter dependence of the data.
\end{abstract}


\end{frontmatter}
\vspace{-7mm} PACS: 12.38.Lg, 11.30, 12.38.-t

\section{Introduction}

Recently, transverse-lattice calculations have provided first data
\cite{Dalley:2003sz} on the {\em impact-parameter dependent} diagonal
(non-skewed) non-singlet generalized parton distributions of the pion.
Generalized parton distribution (GPD) have been a subject of intense
studies in recent years \cite{mul,rad1,ji1,col,pire,freund} (for a review see,
{\em e.g.}, Ref.~\cite{max,diehl} and references therein) providing a
unified framework for numerous high-energy phenomena. The
impact-parameter-space formulation has been pursued in
Refs.~\cite{Burkardt:2000za,Burkardt:2001ni,Burkardt:2002hr,Diehl:2002he,pasza}.
Actually, this is the natural framework for the transverse lattice QCD
formulation~\cite{Burkardt:2001mf,Burkardt:2001jg,Dalley:2002nj,Dalley:2003sz}. In
addition, the diagonal (non-skewed) distributions incorporate
radiative corrections according to the standard DGLAP evolution
equations for not-too-small values of the impact parameter
$b$~\cite{GM,bel}. The results of Ref.~\cite{Dalley:2003sz} may also
provide some guidance on the yet unknown low-$b$ evolution of the
GPD's.

In this paper we obtain theoretical predictions for the GPD from
two different chiral quark models, {\em i.e.}, models incorporating the 
dynamical chiral symmetry breaking: the recently-proposed {\em Spectral
Quark Model}~\cite{RuizArriola:2001rr,RuizArriola:2003bs} and the {\em
Nambu--Jona-Lasinio} model with the Pauli-Villars
regulator~\cite{DR95,WRG99,DR02,RuizArriola:2002wr,TNV}. For these models
it has already been shown that the $b$-integrated (forward) parton
distribution functions agree remarkably well with the phenomenological
parameterization at $Q^2= 4~{\rm GeV}^2$~\cite{SMRS92}. Our very
simple predictions for the GPD, pertaining to a low scale of about
320~MeV, are then evolved with the help of the standard DGLAP
equations to the scales corresponding to the transverse-lattice
calculations~\cite{Burkardt:2001mf,Burkardt:2001jg,Dalley:2002nj,Dalley:2003sz}.
After the evolution the results of Sec.~\ref{sec:results} are in a
good qualitative agreement with the data, showing similar Bjorken-$x$
dependence in the corresponding impact-parameter bins.

\section{Definitions}

The off-forward ($\bf{\Delta_ \perp} \neq 0$) diagonal ($\xi=0$)
generalized parton distribution of the pion is defined
by~\cite{Burkardt:2002hr}\footnote{We drop the quark flavour index
since, {\em e.g.}, for a positively charged pion, $\pi^+$, one has $ H_u (x,
0,t)= H_{\bar d} (1-x,0,t)$.}
\begin{eqnarray}
H(x,\xi = 0, - {\bf{\Delta}}_ \perp^2 ) &=& \int d^2 b \int
\frac{dz^-}{4\pi} e^{i ( x p^+ z^- + {\bf{\Delta}}_\perp \cdot
{\bf{b}} )} \nonumber \\ &\times & \langle \pi^+ (p') | \bar q (0,
-\frac{z^-}{2} , {\bf{b}} ) \gamma^+ q (0, \frac{z^-}{2} , {\bf{b}} )
| \pi^+ (p) \rangle ,
\end{eqnarray} 
where $x$ is the Bjorken $x$, and $\Delta_\perp=p'-p$ lies in the
transverse plane.  This function has the interesting
properties,
\begin{eqnarray}
\int_0^1 dx H(x,0,- {\bf{\Delta}}_\perp^2 ) = F( -
{\bf{\Delta}}_\perp^2 ), \;\;\;\;
H(x,0, - {\bf{\Delta}}_\perp^2=0 ) = q(x) ,
\label{eq:prop} 
\end{eqnarray} 
relating it to the pion electromagnetic form factor, $F(t)$, and to
the pion forward parton distribution, $q(x)$. One can introduce the
impact-parameter representation~\cite{Burkardt:2002hr},
\begin{eqnarray} 
\hspace{-10mm} q ( {\bf b}, x ) = \int \frac{d^2 \Delta_\perp}{(2\pi)^2}
e^{-{\rm i} {\bf b}\cdot {\bf \Delta}_\perp } H(x,0,-{\bf
\Delta}_\perp^2 )= \int_0^{\infty} \frac{\Delta_\perp d
\Delta_\perp}{2\pi} J_0 (b \Delta_\perp) H(x,0,-{\bf \Delta}_\perp^2).
\label{fb}
\end{eqnarray} 
where the cylindrical symmetry has been used.  The second of
Eq.~(\ref{eq:prop}) corresponds to $ \int d^2 b \, q(x,{\bf b}) = q(x) $.

\section{Evaluation in chiral quark models \label{sec:model}}

\begin{figure}[tb]
\begin{center}
\includegraphics[width=7.5cm]{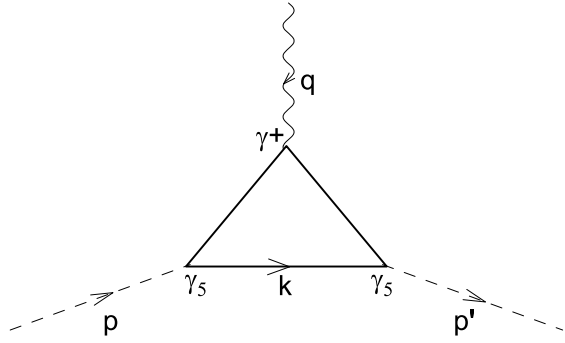}
\end{center} 
\caption{The diagram for the evaluation of the generalized parton distribution of the pion in chiral quark
models.}
\label{fig:diag}
\end{figure}

In chiral quark models the evaluation of $H$ at the leading-$N_c$
(one-loop) level amounts to the calculation of the diagram of
Fig.~\ref{fig:diag},
where the solid line denotes the propagator of the quark of mass
$\omega$. Formally\footnote{The gauge invariant regularizations, allowing
to shift the momentum in the integral, will be specified later.}, this
yields
\begin{eqnarray}
\hspace{-7mm} H(x,0,-{\bf \Delta}_\perp^2; \omega) &=& \frac{i N_c \omega^2}{f_\pi^2} 
\int {d^4 k \over (2\pi)^4 } {\rm Tr} \left[ \gamma^+
{1\over \slashchar{k}-\slashchar{p'} - \omega} \gamma_5 {1\over
\slashchar{k} - \omega} \gamma_5 {1\over
\slashchar{k}-\slashchar{p} - \omega} \right] \nonumber \\
&\times & \delta \left[k^+ - (1-x) p^+ \right], 
\end{eqnarray} 
with $f_\pi=93$~MeV denoting the pion decay constant and $p^2=p'^2=m_\pi^2$. The light-cone coordinates are defined  as
\begin{eqnarray}
k^+ = k^0 + k^3, \; k^- = k^0 - k^3, \; \vec k_\perp =(k^1 ,
k^2 ), \; dk^0 dk^3 = \frac12 dk^+ dk^- .
\end{eqnarray} 
The calculation becomes simplest in the Breit frame, $\Delta^+
=0$. The Cauchy theorem can be applied for the $k^-$
integration~\cite{Frederico:ye}, yielding, after integration and in the subsequent
chiral limit of $m_\pi \to 0$, the result
\begin{eqnarray}
H(x,0,-{\bf \Delta}_\perp^2; \omega) = \frac{N_c \omega^2}{\pi
f_\pi^2} \int \frac{d^2 {\bf K}_\perp}{(2\pi)^2} \frac{\left[ 1 +
\frac{ {\bf K}_\perp \cdot \Delta_\perp (1-x)}{{\bf K}_\perp^2 +
\omega^2} \right]}{({\bf K}_\perp+(1-x)\Delta_\perp)^2 + \omega^2},
\label{eq:unint_ff}
\end{eqnarray} 
where the relative perpendicular momentum is $ {\bf K}_\perp
=(1-x){\bf p}_\perp - x {\bf k}_\perp$.

To proceed further, we need to specify the regularization. First, we
consider the recently proposed {\em Spectral Quark Model}
\cite{RuizArriola:2001rr,RuizArriola:2003bs}. The approach is successful in
describing both the low- and high-energy phenomenology of the pion, and it
complies to the chiral symmetry, including the anomalies. The model amounts to
supplying the quark loop with an integral over $\omega$ weighted by a
quark spectral density $\rho(\omega)$,
\begin{eqnarray}
H(x,0,-{\bf \Delta}_\perp^2) = \int_C d\omega\,\rho(\omega) H(x,0,-{\bf \Delta}_\perp^2; \omega),
\label{eq:unint_spect}
\end{eqnarray} 
where $C$ is a suitably chosen integration contour in the complex
$\omega$ space \cite{RuizArriola:2003bs}.  Next, we apply the simple techniques
described in detail in Ref.~\cite{RuizArriola:2003bs}, use the Feynman
trick for the two denominators in Eq.~(\ref{eq:unint_ff}), and
integrate over ${\bf K}_\perp$. The result is
\begin{eqnarray}
\!\!\!\!\!\!\! H(x,0,-{\bf \Delta}_\perp^2) = 1+\frac{N_c}{8 \pi^2 f_\pi^2} \int
\omega^2 \rho(\omega) d \omega \int_0^1 d \alpha \frac{(1-x)^2 {\bf
\Delta}_\perp^2 }{\omega^2 + \alpha (1-\alpha) (1-x)^2 {\bf
\Delta}_\perp^2 )}.
\label{eq:unint_ff5}
\end{eqnarray} 
Note the correct normalization condition, $
F(0)=1$. Moreover, the pion electromagnetic ms radius is 
$\langle r^2 \rangle \equiv -6 dF(t)/dt|_{t=0} = N_c/({4 \pi^2 f_\pi^2})$. 

In the {\em Meson Dominance} variant \cite{RuizArriola:2003bs} of the Spectral Quark Model 
the relevant part of the spectral function has the form
\begin{eqnarray}
\rho_V (\omega) = \frac{1}{2\pi i} \frac{3 \pi^2 m_\rho^3 f_\pi^2 }{4 N_c}
\frac{1}{\omega} \frac1{(m_\rho^2/4-\omega^2)^{5/2}}, 
\label{v52}
\end{eqnarray} 
where $m_\rho=770$~MeV is the
mass of the $\rho$ meson\footnote{In this case the relation $ m_\rho^2
= 24 \pi^2 f_\pi^2 /N_c$ holds \cite{RuizArriola:2003bs}.}.  The function $\rho_V(\omega)$ has a single
pole at the origin and branch cuts starting at $\pm m_\rho/2$.  The
contour $C$ encircles the branch cuts, {\em i.e.}, starts at
$-\infty+i0$, goes around the branch point at $-m_\rho/2$, and returns
to $-\infty -i0$, with the other section obtained by a reflexion with respect
to the origin%
~\cite{RuizArriola:2003bs}. In the Meson Dominance model we get
from (\ref{eq:unint_ff}) and (\ref{v52}) the explicit result of an
appealing simplicity, namely
\begin{eqnarray} 
H(x,0,- {\bf \Delta}_\perp^2) = \frac{m_\rho^2 ( m_\rho^2 - (1-x)^2
{\bf \Delta}_\perp^2)} {( m_\rho^2 + (1-x)^2 {\bf
\Delta}_\perp^2)^2}. \label{res:vmd}
\end{eqnarray} 
We check that $H(x,0,0)=1 $~\cite{RuizArriola:2003bs} and $\int_0^1 dx
H(x,0,t) = m_\rho^2/(m_\rho^2+t)$, Eq.~(\ref{eq:prop}), which is the
built-in vector-meson dominance principle. We pass to the
impact-parameter space by the Fourier-Bessel transformation (\ref{fb})
and get
\begin{eqnarray} 
q ( {\bf b}, x ) = \frac{m_\rho^2 }{ 2\pi (1-x)^2 }\left[ K_0 \left( \frac{b
m_\rho}{1-x} \right) - \frac{b m_\rho}{1-x} K_1 \left( \frac{b
m_\rho}{1-x} \right) \right].
\label{k0k1}
\end{eqnarray} 

In the Nambu--Jona-Lasinio model with the Pauli-Villars regularization one can
proceed along similar lines as above to get
\begin{eqnarray}
H(x,0,- {\bf \Delta}_\perp^2) &=& 1-\frac{N_c M^2}{8 \pi^2 f_\pi^2}
\sum_i c_i \int_0^1 d \alpha \frac{(1-x)^2 {\bf \Delta}_\perp^2 }{M^2
+ \Lambda_i^2 + \alpha (1-\alpha) (1-x)^2 {\bf \Delta}_\perp^2
)}\nonumber \\ &=&1+\frac{N_c M^2 (1-x)|{\bf \Delta}_\perp|}{4 \pi^2
f_\pi^2 s_i} \sum_i c_i \log \left ( \frac{s_i+(1-x)|{\bf
\Delta}_\perp|}{s_i-(1-x)|{\bf \Delta}_\perp|} \right ), \nonumber \\
s_i&=&\sqrt{(1-x)^2{\bf \Delta}_\perp^2+4M^2+4\Lambda_i^2} ,
\label{eq:unint_ff4}
\end{eqnarray} 
where $M$ is the constituent quark mass, $\Lambda_i$ are the PV
regulators, and $c_i$ are suitable constants.  For the
twice-subtracted case, explored below, one has, for any regulated
function $F$, the operational definition~\cite{RuizArriola:2002wr}
\begin{eqnarray}
\sum_i c_i F( \Lambda_i^2 ) = F(0) - F(\Lambda^2 ) + \Lambda^2
dF(\Lambda^2 )/d\Lambda^2.
\label{eq:PV2} 
\end{eqnarray} 
In what follows we use $M=280$~MeV and $\Lambda=871$~MeV, which yields
$f_\pi=93.3$~MeV \cite{RuizArriola:2002wr}.

It is interesting to notice that, quite generally, the chiral quark model 
results displayed above depend on the momentum ${\bf \Delta}_\perp$ and $x$ 
only through the combination $(1-x)^2{\bf \Delta}_\perp^2$. Consequently, 
in the $b$ space they depend on the combination $b^2/(1-x)^2$.
Due to this property we have 
\begin{eqnarray} 
\frac{\int d^2b \, b^{2n} q(b,x) }{\int d^2b \,q(b,x) } \equiv
\langle b^{2n} \rangle (x) = (1-x)^{2n} \langle b^{2n} \rangle (0).
\end{eqnarray} 
This means, that all the moments except for $n=0$ vanish as $x \to 1$,
or, in other words, the function becomes an infinitely-narrow $\delta$
function in this limit.  This general prediction of chiral quark
models is clearly seen in the lattice data of Ref.~\cite{Dalley:2003sz},
{\em cf.} Fig.~2(b).

\section{Smearing over $b$}

\begin{figure}[tb]
\begin{center}
\includegraphics[width=9cm]{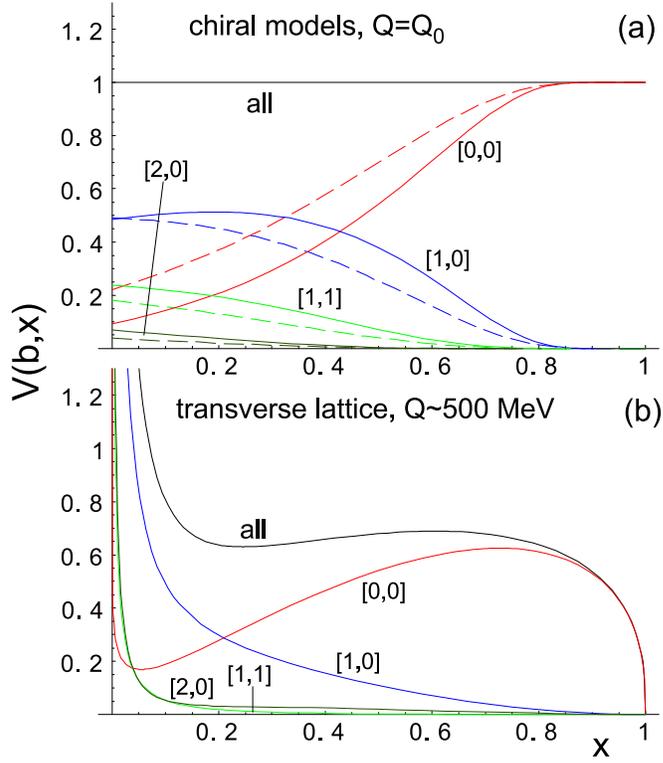}
\end{center} 
\caption{Valence impact-parameter dependent diagonal 
GPD of the pion, $V(x,b)$, plotted as a function of the Bjorken $x$ variable.
(a) The results of the chiral quark models at the model scale of $Q=Q_0=313$~MeV. Solid lines: the
Spectral Quark Model of Ref.~\cite{RuizArriola:2003bs}, dashed lines: the
Nambu--Jona-Lasinio model with two Pauli-Villars subtractions. Label
{\em all} denotes the forward distribution, {\em i.e.}, the function
$V(x,b)$ integrated over the whole $b$-plane. Labels $[i,j]$ denote
the function $V(x,b)$ integrated over the square plaquettes centered
at coordinates $(i b_0,j b_0)$ of the edge of length $b_0$, times the
degeneracy of the plaquette (see the text for details). Following
Ref.~\cite{Dalley:2003sz}, the value of $b_0$ is taken to be $2/3$~fm.
(b) The results for $V(x,b)$ at the scale $Q \sim 500$~MeV, obtained
from transverse-lattice calculation of
Ref.~\cite{Dalley:2003sz}. Labels as in (a). The model results of (a) can be compared to the data of
(b) only after a suitable QCD evolution.}
\label{fig:fig2}
\end{figure}

Our aim is to compare our results, after a suitable QCD evolution,
to the transverse-lattice data of Ref.~\cite{Dalley:2003sz}. These
data give the non-singlet diagonal parton distribution of the pion at
discrete values of the impact parameter ${\bf b}$, corresponding to a
square lattice with spacing of $b_0\simeq 2/3$~fm. 
It is certainly not obvious how to compare discrete data to a continuum model. Clearly, we 
cannot achieve the continuum limit on transverse lattices, on the other hand we do not intend, 
in a simple study as presented here, to put chiral models on the lattice. 
A simple and reasonable
comparison \cite{dalley:priv} is expected when the model predictions are {\em smeared over
square plaquettes}, the same ones as in the discrete lattice. The plaquettes are labeled $[i,j]$, 
which means that they are centered at coordinates
$(i b_0,j b_0)$, and have the edge of length $b_0=2/3$~fm \cite{Dalley:2003sz}. The smeared 
GPD is defined as
\begin{eqnarray}
V(x,[i,j])\equiv \int_{(i-1/2) b_0}^{(i+1/2) b_0} db_1 \int_{(j-1/2) b_0}^{(j+1/2) b_0} db_2 
V(x,\sqrt{b_1^2+b_2^2}).
\label{smear} 
\end{eqnarray}
Figure \ref{fig:fig2} shows the results of this smearing. In addition, the degeneracy factor of
the number of plaquettes 
equidistant from the origin is included, {\em i.e.}, the $[1,0]$, $[1,1]$, and $[2,0]$ plaquettes are multiplied 
by a factor of four, while $[2,1]$ would be multiplied by eight. 

We note that the smearing has a large effect for the $[0,0]$
plaquette. This is because in the limit of $x \to 1$ the function
$V(x,b)$ becomes a distribution in $b$, which can be seen immediately
from the explicit form of Eq.~(\ref{k0k1}). Thus, the results for
$[0,0]$ are sensitive to the size of $b_0$. For lower values of $b_0$
the function becomes very sharply peaked at $x=1$.

Figure 2(b) shows the data from the transverse-lattice calculations
shown by Dalley in Ref.~\cite{Dalley:2003sz}.  These data correspond
to the scale $Q \simeq 500$~MeV, as inferred in Ref.~\cite{Dalley:2002nj} 
from the analysis of the pion light-cone wave function.
Since the scale pertaining to our calculation is much lower, we
need to evolve our results upward before comparing to the data of Fig.~2(b).

\section{QCD evolution \label{sec:evol}} 
The simple calculation of Sec.~\ref{sec:model} has produced distributions
corresponding to a low quark model scale, $Q_0$. A priori, the value
of $Q_0$ is not known. The way to estimate it is to run the QCD
evolution upward from various scales $Q_0$ up to a scale $Q$ where the
data can be used.  Alternatively, one may use the
momentum fraction carried by the quarks at the scale $Q$ and the
downward QCD evolution in order to estimate $Q_0$~\cite{DR95,DR02,RuizArriola:2002wr}.
We use the LO evolution with 
\begin{eqnarray} 
 \alpha(Q)= \left( \frac{4 \pi}{\beta_0 } \right) \frac1{\log (Q^2
 / \Lambda_{\rm QCD}^2 )},
\end{eqnarray} 
where $ \beta_0 = 11 C_A /3 - 2 N_F /3 $, $C_A = 3$, and $N_F=3$ is the
number of active flavors. We take $\Lambda_{\rm QCD}=226~{\rm MeV} $, which for 
$Q = 2~{\rm GeV}$ yields $\alpha=0.32 $~\cite{PDG}.
Then one proceeds as follows:
The valence contribution to the energy momentum tensor evolves as 
the first $x$-moment of the valence quark distribution, 
\begin{eqnarray}
\frac{ V_1 (Q) } { V_1 (Q_0) } = \left( \frac{\alpha(Q)}
{\alpha(Q_0) } \right)^{\gamma_1^{\rm NS} / (2 \beta_0) } ,
\end{eqnarray} 
where $ \gamma_1^{\rm NS} / (2 \beta_0) = 32/81$.  The
value of $V_1 (Q)$ has been extracted from the analysis of 
high-energy experiments. In Ref.~\cite{SMRS92} it was found that at 
\mbox{$Q=2$~GeV} the valence quarks carry $47 \%$ of the total momentum
of the pion, {\em e.g.}, for $\pi^+$
\begin{eqnarray} 
V_1 = \langle x \left( u_\pi - \bar u_\pi + \bar d_\pi - d_\pi \right)
\rangle = 0.47 \pm 0.02 \qquad {\rm at} \qquad Q = 2~{\rm GeV}.
\end{eqnarray} 
The downward LO DGLAP evolution yields at the scale $Q_0$
\begin{eqnarray}
V_1 (Q_0) = 1, \;\;\; G_1 (Q_0) + S_1 (Q_0) = 0.
\end{eqnarray} 
with $G_1$ and $S_1$ the gluon and sea momentum fractions,
respectively.  The scale $Q_0$ defined with this prescription is
called the {\em quark model point}, since obviously in effective quark
models all the momentum is carried by the quarks.  At LO the scale
turns out to be \cite{DR95}
\begin{eqnarray}
Q_0 = 313_{-10}^{+20}~{\rm MeV}.  
\label{eq:mu0_dis} 
\end{eqnarray} 
This is admittedly a rather low scale, but one can still hope that the
typical expansion parameter $\alpha (Q_0)/(2 \pi) \sim 0.34 \pm 0.04 $
makes the perturbation theory meaningful. Actually, the NLO analysis of
Ref.~\cite{DR02} supports this assumption. In addition, this is the
same scale used in Ref.~\cite{RuizArriola:2002bp} to compute the pion
LC wave function. 
\footnote{An analogous analysis applied to the data of Ref.~\cite{Dalley:2003sz}
shows that the momentum fraction carried by the valence quarks is $72\%$ \cite{dalley:priv}, 
which at LO would imply the scale of 477~MeV, compatible 
with the scale of $500$~MeV quoted by the authors of Ref. \cite{Dalley:2003sz}.}

Following Ref.~\cite{GM}, we apply the DGLAP evolution to the
off-forward diagonal distribution function with the evolution kernel
that does not depend on ${\bf \Delta}_\perp$, or, in the
impact-parameter space, on $b$. 
Then, at LO the DGLAP
evolution in the index space simply reads
\begin{eqnarray}
\!\!\!\! V_n (Q,b) \equiv \int_0^1 dx \, x^{n} V(x,Q,b) = \left(
\frac{\alpha(Q)} {\alpha(Q_0) } \right)^{\gamma_{n}^{\rm NS} / (2
\beta_0) } \int_0^1 dx \, x^{n} V(x,Q_0,b),
\label{eq:evol_val} 
\end{eqnarray} 
where the anomalous dimension is 
\begin{eqnarray}
\gamma_n^{\rm NS} &=& -2 C_F \left[ 3 + \frac{2}{(n+1)(n+2)}- 4
\sum_{k=1}^{n+1} \frac1k \right],
\label{eq:anom_dim} 
\end{eqnarray}
with $C_F = 4/3$. With $n$ treated as a complex number, which requires
an analytic continuation of both $V_n (Q_0,b) $ and $\gamma_n^{\rm
NS}$, Eq.~(\ref{eq:evol_val}) can be inverted using the inverse Mellin
transform
\begin{equation}
V (x,Q,b)=\int_{-i\infty}^{+i\infty}{{\rm d} n\over{2\pi {i}}}x^{-n-1}
V_n (Q,b).
\label{eq:inv_mellin} 
\end{equation}
The procedure, carried out numerically, is fast and stable. Since the
singularity structure of $V_n (Q,b)$ is the same as for the forward
case, we may use the standard Mellin integration contour in
Eq.~(\ref{eq:inv_mellin}).

An interesting feature of the above evolution is the induced
suppression at \mbox{$x \to 1$}. Thus, using known methods from the
$b-$ integrated case \cite{Peterman:1978tb}, a function which
originally behaves as \mbox{$V(x,Q_0,b) \to C(b) (1-x)^N$} evolves
into
\begin{eqnarray} 
V (x,Q,b) \to C(b) (1-x)^{N - \frac{4 C_F }{\beta_0} \log
\frac{\alpha(Q)}{ \alpha(Q_0) }}, \qquad x\to 1.
\label{endpoint}
\end{eqnarray} 

In order to compare to the transverse-lattice data of
Ref.~\cite{Dalley:2003sz}, we apply the evolution to the smeared
functions of Eq.~(\ref{smear}). Thus, we have explicitly
\begin{eqnarray}
\!\!\!\!\!\! V (x,Q,[i,j])=\int_{-i\infty}^{+i\infty}{{\rm d} n\over{2\pi
{i}}}x^{-n-1} \left( \frac{\alpha(Q)} {\alpha(Q_0) }
\right)^{\gamma_{n}^{\rm NS} / (2 \beta_0) } \int_0^1 dy \, y^{n}
V(y,Q_0,[i,j]),
\label{ijevol}
\end{eqnarray}
where the distribution at the scale $Q_0$ is the prediction of either
of the two considered chiral quark models.

We also note that in the Spectral Quark Model 
\begin{eqnarray}
&&\hspace{-10mm}V_n(Q_0,b)= \\ &&\hspace{-7mm}\frac{m_\rho^2\Gamma
(n+1)}{\pi 2^{n+3}} \left [ b m_\rho G_{2,4}^{4,0}\left .
\Bigg(\frac{{b^2} {m_\rho^2}}{4}\right | \begin{array}[c]{c}
\frac{n-1}{2},\frac{n}{2} \\ -1,-\frac{1}{2},-\frac{1}{2},\frac{1}{2}
\end{array}\Bigg)-G_{2,4}^{4,0}\left .
\Bigg(\frac{{b^2}{m_\rho^2}}{4}\right | \begin{array}[c]{c}
\frac{n}{2},\frac{n+1}{2} \\ -\frac{1}{2},0,0,0
\end{array}\Bigg)\right ], \nonumber
\end{eqnarray}
where $G$ denotes the Meijer $G$ function. This form can be useful for
further analytic considerations.

\section{Results and conclusions \label{sec:results}}

Figure 2 (a) shows the plaquette-averaged functions $V(x,Q_0,[i,j])$
for the Spectral Quark Model (solid lines) and the NJL model (dashed
lines). We note that the predictions of the two models are
qualitatively the same, with the NJL curves pushed to somewhat lower
values of $x$. For the lack of space, in this paper we display the QCD evolution
of the Spectral Quark Model only. The case of the NJL model 
is qualitatively the same, with the corresponding curves moved to 
a bit lower values of $x$, 
simply reflecting the different initial condition of Fig. 2 (a).
These results and other details will be presented in a longer paper.

The results of the evolution are shown in Fig.~3 at three
values of the reference scale $Q$: 400~MeV (a), 500~MeV (b), and 2~GeV
(c).  We note a large effect of the evolution on the distribution
functions. The lines labeled {\em all} correspond to the forward case,
{\em i.e.}, show 
\mbox{$\int d^2b \, V(x,Q,b)=V(x,Q,{\bf \Delta}_\perp=0)$}. The
originally flat distribution of Fig.~2(a) recovers the correct
end-point behavior at $x \to 1$ according to Eq.~(\ref{endpoint}). As
$Q$ increases, the distribution is pushed towards lower values of $x$,
as is well known for the DGLAP evolution. At $Q=2$~GeV the result
agrees very well with the SMRS parameterization of the pion structure
function \cite{SMRS92}, as can be seen from Fig.~3(d) 
(here we plot  for convenience $x V(x,Q)$) by comparing the
dashed and solid lines. This result was already obtained in
Refs.~\cite{DR95,DR02}.

\begin{figure}[tb]
\begin{center}
\includegraphics[width=14cm]{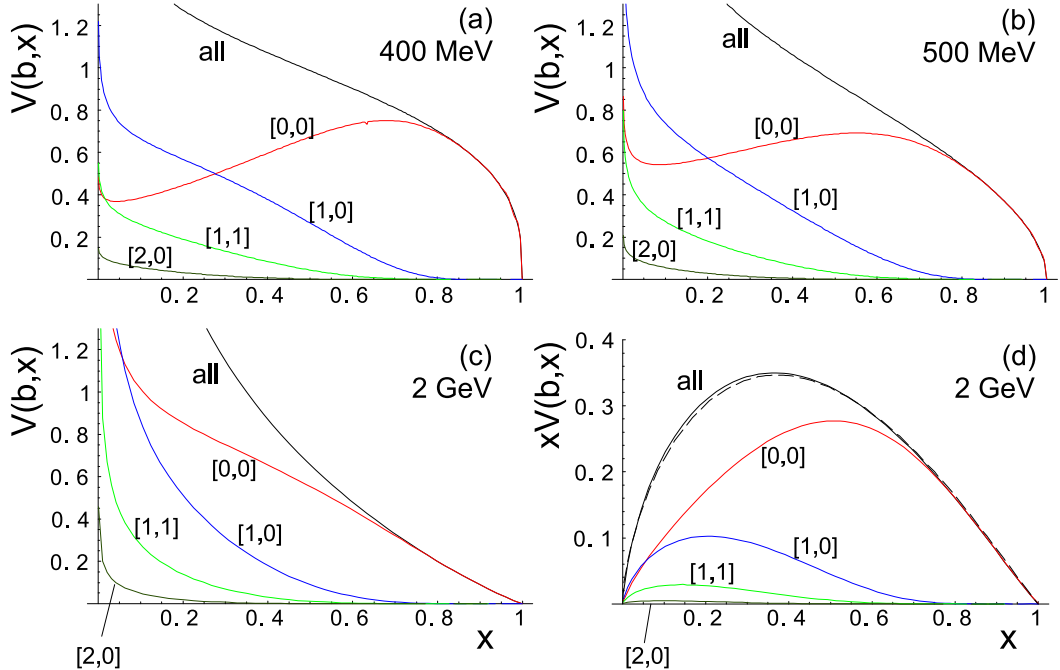}
\end{center} 
\caption{Results of the LO DGLAP evolution of the impact-parameter
dependent diagonal non-singlet generalized parton distribution function of the
pion, $V(x,b,[i,j])$, started from the initial condition at
$Q=Q_0=313$~MeV produced by the Spectral Quark Model (Fig.~2(a), solid
lines). Figures (a,b,c) correspond to $Q=400$~MeV, $500$~MeV, and
$2$~GeV, respectively. Labels as in Fig.~2. Figure (d) shows $x
V(x,b,[i,j])$ for $Q=2$~GeV, with the dashed line showing the SMRS
\cite{SMRS92} parameterization of the data for the forward parton distribution function.
}
\label{fig:evo}
\end{figure}

The results for the plaquette $[0,0]$ follow, at large $x$, the
forward distributions. This is clear from the behavior described at
the end of Sec.~\ref{sec:model}, {\em i.e.}, from the dependence of the
initial function on the variable $b/(1-x)$. Certainly, as $x \to 1$,
the integration over the $[0,0]$ plaquette is the same as the
integration over the whole $b$-space. At $Q=400$~MeV and $500$~MeV the
values of $V(x,Q,[0,0])$ reach a maximum at an intermediate value of
$x$, and develop a dip at low $x$. This is in qualitative agreement
with the transverse-lattice data of Fig.~2(b). We note that there the dip at
low $x$ is lower than in our model calculation, yet, in view of the
simple nature of our model and approximations (chiral limit, LO
evolution, evolution independent of $b$, uncertainties in the
determination of $b_0$ and $Q$ on the lattice) the similarity is quite
satisfactory. We have checked that if the value the lattice-spacing
parameter, $b_0$, were lowered, an even more quantitative agreement
would follow.

The results for non-central plaquettes also qualitatively agree with
the lattice measurements.  In this case at $x \to 1$ the corresponding
functions vanish very fast, in accordance to our model formulas.  The
difference with the lattice calculation of Fig.~2(b) is that in our
case the farther plaquettes naturally bring less and less, and the
yield from the $[2,0]$ plaquette is lower than for the $[1,1]$
plaquette. In Fig.~2(b) it is the other way around.

In summary, the obtained agreement of our approach, based on non
perturbative chiral quark models in conjunction with perturbative LO
DGLAP evolution, with the data from the transverse lattices, is quite
remarkable and encouraging, baring in mind the simplicity of the
models and the apparently radically different handling of chiral
symmetry in both approaches. We also note that the low-energy scale
taken for the chiral quark models is consistent with previous analysis
based both on the forward parton distribution amplitudes as well as
the light cone wave function. Our analysis might be reinforced by
extending our calculation to include the NLO perturbative
corrections. Such a study is left for a future research. 


\section*{Acknowledgements}
We are grateful to Simon Dalley for helpful discussions concerning the
transverse-lattice data, and to Krzysztof Golec-Biernat for a
discussion on the validity of the DGLAP evolution for the
impact-parameter dependent GPD's.


\begin{thebibliography}{99}

\bibitem{Dalley:2003sz}
S.~Dalley, {\em Impact parameter dependent quark distribution of the pion},
hep-ph/0306121.

\bibitem{mul} D. M\"uller, D. Robaschik, B. Geyer, F.-M. Dittes, and
J. Horejsi, Fortschr. Phys. 42 (1994) 101.

\bibitem{rad1} A. V. Radyushkin, Phys. Lett. B380 (1996) 417;
Phys. Lett. B385 (1996) 333; Phys. Rev. D56 (1997) 5524.

\bibitem{ji1} X. Ji, Phys. Rev. Lett. 78 (1997) 610; Phys. Rev. D55
(1997) 7114; J. Phys. G 24 (1998) 1182.

\bibitem{col} J. C. Collins, L. Frankfurt, and M. Strikman,
Phys. Rev. D56 (1997) 2982.

\bibitem{pire} R. Ralston and B. Pire, Phys. Rev. D66 (2002) 111501.

\bibitem{freund} A. Freund, hep-ph/0212017, hep-ph/030612.

\bibitem{max} K. Goeke, M. V. Polyakov, and M. Vanderhaeghen,
Prog. Part. Nucl. Phys. 47 (2001) 401.

\bibitem{diehl} M. Diehl, {\em Generalized Parton Distributions}, DESY-THESIS-2003-018,
hep-ph/0307382.

\bibitem{Burkardt:2000za}
M.~Burkardt,
Phys.\ Rev.\ D {\bf 62} (2000) 071503
[Erratum-ibid.\ D {\bf 66} (2002) 119903].


\bibitem{Burkardt:2001ni} 
M.~Burkardt,
invited talk at Workshop on Lepton Scattering,
Hadrons and QCD, Adelaide, Australia, 26 Mar - 6 Apr 2001, hep-ph/0105324. 

\bibitem{Burkardt:2002hr}
M.~Burkardt,
Int.\ J.\ Mod.\ Phys.\ A {\bf 18} (2003) 173.

\bibitem{Diehl:2002he}
M.~Diehl,
Eur.\ Phys.\ J.\ C {\bf 25} (2002) 223.


\bibitem{pasza} P. V. Pobylitsa, Phys. Rev. D65 (2002) 077504; Phys. Rev. D65 (2002) 114015.

\bibitem{Burkardt:2001mf}
M.~Burkardt and S.~K.~Seal,
Phys.\ Rev.\ D {\bf 65} (2002) 034501.


\bibitem{Burkardt:2001jg}
M.~Burkardt and S.~Dalley,
Prog.\ Part.\ Nucl.\ Phys.\  {\bf 48} (2002) 317.

\bibitem{Dalley:2002nj}
S.~Dalley and B.~van de Sande,
Phys.\ Rev.\ D {\bf 67} (2003) 114507.

\bibitem{GM} K. J. Golec-Biernat and Alan D. Martin, Phys. Rev. D59
(1999) 014029.

\bibitem{bel} A.~V.~Belitsky, D.~M\"uller, and A.~Kirchner,
Nucl.\ Phys.\ B {\bf 629}, 323 (2002).

\bibitem{RuizArriola:2001rr}
E.~Ruiz Arriola,
talk given at Workshop on Lepton Scattering, Hadrons and QCD, Adelaide, Australia, 26 Mar - 6 Apr 2001,
hep-ph/0107087.

\bibitem{RuizArriola:2003bs}
E.~Ruiz Arriola and W.~Broniowski,
Phys.\ Rev.\ D {\bf 67}, 074021 (2003)

\bibitem{DR95} R. M. Davidson and E. Ruiz Arriola, Phys. lett. {\bf B
348} (1995) 163.

\bibitem{WRG99} H. Weigel, E. Ruiz Arriola, and L. P. Gamberg,
Nucl. Phys. {\bf B 560} (1999) 383.

\bibitem{DR02} R. M. Davidson and E. Ruiz Arriola,
Act. Phys. Pol. {\bf B 33} (2002) 1791. 

\bibitem{RuizArriola:2002wr}
E.~Ruiz Arriola,
lectures given at 42nd
Cracow School of Theoretical Physics, {\em Flavor
Dynamics}, Zakopane, Poland, 31 May - 9 Jun 2002, Acta Phys.\ Polon.\ B {\bf 33} (2002) 4443. 

\bibitem{TNV} L.~Theu$\beta$el , S. Noguera and V. Vento,
nucl-th/0211036.

\bibitem{SMRS92} P. J. Sutton, A. D. Martin, R. G. Roberts, and W.J.
Stirling, Phys. Rev. {\bf D 45} (1992) 2349.

\bibitem{Frederico:ye}
T.~Frederico and G.~A.~Miller,
Phys.\ Rev.\ D {\bf 45} (1992) 4207.

\bibitem{dalley:priv} S. Dalley, private communication.

\bibitem{PDG} Review of Particle Physics, K. Hagiwara {\em et al.},
Phys. Rev. {\bf D 66} (2002) 010001. 

\bibitem{RuizArriola:2002bp}
E.~Ruiz Arriola and W.~Broniowski,
Phys.\ Rev.\ D {\bf 66}, 094016 (2002).


\bibitem{Peterman:1978tb}
A.~Peterman,
Phys.\ Rept.\  {\bf 53} (1979) 157.

\end{thebibliography}
\end{document}